\documentstyle[a4,12pt,epsf]{article}

\newcommand{\gev}{{\rm GeV}}
\newcommand{\fm}{{\rm fm}}

\newcommand{\eqn}[1]{Equation~(\ref{#1})}

\newcommand{\fig}[1]{Figure~(\ref{#1})}
\newcommand{\tab}[1]{Table~(\ref{#1})}
\newcommand{\refer}[1]{Reference~\cite{#1}}

\begin{document}
\begin{titlepage}
\begin{flushright}
Edinburgh Preprint: 95/559\\
Liverpool Preprint: LTH-363\\
hep-lat/9511031
\end{flushright}
\vspace*{5mm}

\begin{center}
{\Huge The Landshoff-Nachtmann Pomeron on the Lattice}\\[15mm]
{\large\it UKQCD Collaboration}\\[3mm]

{\bf D.S.~Henty, C.~Parrinello\footnote{Present address: D.A.M.T.P.,
University of Liverpool, Liverpool L69 3BX, U.K.} and D.G.~Richards
}\\ Department of Physics and Astronomy, University of Edinburgh,
Edinburgh EH9 3JZ, Scotland\\[2.0ex]

\end{center}
\vspace{5mm}

\begin{abstract}
We investigate the Landshoff-Nachtmann two-gluon-exchange
model of the Pomeron using gluon propagators computed in the
Landau gauge within quenched lattice QCD calculations. We first
determine an effective gluon-quark coupling by constraining the
Pomeron-quark coupling to its phenomenological value $\beta_0 = 2\,
\gev^{-1}$.  We then provide predictions for a variety of diffractive
processes. As the propagators have been evaluated entirely from QCD
first principles (although in the quenched approximation), our results
provide a consistency check of the Landshoff-Nachtmann model.  We
address the issue of the possible gauge-dependence of our results,
which will be the object of a future study. \end{abstract}
\end{titlepage}

\section{Introduction}
The description of diffractive hadronic physics in terms of the
exchange of a ``Pomeron'' has proved remarkably durable, and has
successfully withstood the advent of QCD. Interest in the Pomeron has
been renewed by recent results from HERA.

Total cross sections and diffractive processes are essentially soft in
nature, and thus do not fall within the realm of perturbative QCD.
Thus any fixed-order perturbative calculation must be regarded purely
as a model of the interaction, but one which may be representative of
an all-orders QCD result.  Nevertheless, a successful phenomenological
model of Pomeron exchange is provided by two-gluon-exchange
(2GE)~\cite{low:75,nussinov:75}.  Crucial to this model is that the
gluons be non-perturbative, i.e. infra-red finite, both to avoid the
Coulomb singularity at $t = 0$, and to provide the correct
$t$-dependence of the differential cross section~\cite{dgr:85,ln:87}.

Recently there has been a series of papers~\cite{halzen:93,ducati:93}
investigating the phenomenology of 
the 2GE model in the form proposed by 
Landshoff and Nachtmann (LN)~\cite{ln:87}. 
In these papers a  
non-perturbative gluon propagator extracted from the solution of the
Schwinger-Dyson equation~\cite{cornwall:82} is inserted in the LN model.  
Here we will
explore the phenomenology of the same model, using gluon propagators
extracted from lattice calculations in the Landau gauge.

Our method has the advantage that one is inserting in the model 
a genuine, nonperturbative QCD quantity, which has been computed from first 
principles rather than derived from an approximate equation. Also, as we will 
explain in the following, we have only one free parameter in our approach, 
whereas the Schwinger-Dyson solution depends on two parameters. 
For these reasons, we think that our approach provides in principle 
the best chance to test the success of the LN model from the point of view of 
QCD. 

The layout of the remainder of this paper is as follows.  In the next
section we will summarise the lattice calculations, and present the 
results for the lattice gluon propagator.  In section 3 we will
describe how we apply the measured gluon propagator to the calculation
of cross-sections, and compare our results to the experimental data.
In section 4 we comment on the issue of gauge dependence of our results.
Finally we present our conclusions.

\section{Computational details}
We use the standard lattice definition of the gluon fields in terms of
the link variables \cite{mandula:87},
\begin{equation}
A_{\mu}(x) = \frac{U_{\mu}(x) - U^\dagger_{\mu}(x)}{2i a g_0} - \frac{1}{3} 
{\rm Tr} \left( \frac{U_{\mu}(x) - U^\dagger_{\mu}(x)}{2i a g_0} \right),
\end{equation}
where $a$ is the lattice spacing and $g_0$ the bare coupling constant.
The gauge configurations are fixed to the Landau gauge by imposing the
gauge-fixing condition
\begin{equation}
\Delta(x) = \sum_\mu A_\mu(x+\hat{e}_\mu) - A_\mu(x) = 0.
\end{equation}
The accurate implementation of this step is crucial, and we use the
Fourier-accelerated algorithm of reference~\cite{davies:87}.

The unrenormalised gluon propagator in momentum space 
is obtained from the gauge-fixed
gluon fields by taking the Fourier transform
\begin{equation}
D_{\mu\nu}(p) = \langle A_\mu(p) A_\nu(-p)\rangle.
\end{equation}
In the continuum limit, we can write the Landau-gauge propagator 
in the form
\begin{equation}
D_{\mu\nu}(p) = \left\{ g_{\mu\nu} - \frac{p_\mu p_\nu}{p^2}\right\}
D_{\rm lat} (p).\label{eqn:lorentz}
\end{equation}

We performed calculations on two sets of hypercubic lattices, corresponding to 
different physical volumes and lattice spacings. In addition, we used data 
for the gluon propagator as evaluated in \refer{stella:94}; such data are 
particularly useful for an analysis of finite volume effects in our 
calculations. 
The parameters used are 
listed in Table~\ref{tab:params}.  Full details of the method and of the
simulation for the hypercubic lattices are contained in ref.~\cite{dsh_cp:95}.
\begin{table}
\begin{center}
\begin{tabular}{|c|c|c|c|c|}\hline
&$\beta$ & Size & No. Cfgs. & $a^{-1}$ (GeV) \\ \hline
This work & 6.0 & $16^4$ & 150 &  1.9 \\ \hline
This work & 6.2 & $16^4$ & 150 &  2.7 \\ \hline
\refer{stella:94} & 6.0 & $24^3 \times 48$ & 500 & 1.9 \\ \hline
\end{tabular}
\caption{Parameters of the lattices used in the calculation.  
The quoted value of the lattice spacing is obtained from an
analysis of the string tension.}\label{tab:params}
\end{center}
\end{table}

As is customary in lattice QCD, we fix the value of the lattice spacing 
in units of energy from string tension measurements. This sets the  
scale of momenta for the gluon propagator.

\begin{figure}
\begin{center}
\leavevmode
\epsfysize=250pt
\epsfbox[20 30 620 600]{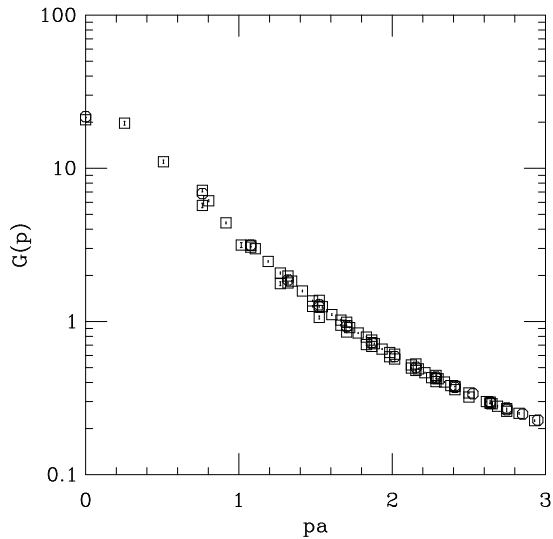}
\end{center}
\caption{The gluon propagator on a $16^4$ lattice at $\beta = 6.0$.}
\label{fig:gluon_prop}
\end{figure}

A representative propagator, $D_{\rm lat} (p)$, is shown in
\fig{fig:gluon_prop}. 
It is important to keep in mind that by defining the theory on a
finite lattice we introduce both an ultraviolet and an infrared cutoff
for Greens functions.  For this reason, our gluon propagator is
infrared finite by construction.  On the other hand, when the lattice
momenta are such that $|pa| \geq 1$, the theory becomes sensitive to
the ultraviolet cutoff and one has to worry about finite lattice
spacing effects.  It is therefore crucial to probe the sensitivity of
our results to the values of the cutoff scales by changing the lattice
parameters as discussed above.

We fit the propagator to the form
\begin{equation}
D_{\rm lat} (p) = \left\{ \begin{array}{ll}
        a_0/(a_1^2 + {p^2}^{1 + a_2}) & |pa| \leq 1 \\
        b_0/({p^2}^{1+b_2}) & |pa| \geq 1
        \end{array}. \right. \label{eq:gluon_form}
\end{equation}
The parameters of the fits in lattice units for each of our hypercubic 
lattices are shown
in Table~\ref{tab:fit_params}; for the remaining lattice, we use the
parameters quoted in \refer{stella:94}.
\begin{table}
\begin{center}
\begin{tabular}{|c|c||c|c|c||c|c|}\hline
$\beta$ & Size & $a_0$ & $a_1$ & $a_2$ & $b_0$ & $b_1$ \\\hline
6.0 & $16^4$ & 3.37(1) & 0.176(1) & 0.445(5) & 3.139(4) & 0.193(1)\\\hline
6.2 & $16^4$ & 2.72(1) & 0.122(1) & 0.461(5) & 2.645(4) & 0.133(1)\\\hline
\end{tabular}
\caption{Parameters of the fit to the gluon propagator on the 
hypercubic lattices, in units of
the lattice spacing, as discussed in the text}\label{tab:fit_params}
\end{center}
\end{table}

\section{Results}
\subsection{Determination of an effective quark-gluon coupling}
 The salient feature of the LN model is
that Pomeron exchange between quarks behaves like a 
$C=+1$ photon-exchange
diagram, with amplitude
\begin{equation}
i \beta^2_0 (\bar{u} \gamma_\mu u)(\bar{u} \gamma^\mu u).
\end{equation}
$\beta_0$ represents the strength of the Pomeron coupling to quarks,
and is related to the (non-perturbative) gluon propagator by
\begin{equation}
\beta_0^2 = \frac{1}{36 \pi^2} \int d^2 p \left[ g^2 D(p)\right]^2,
\label{eq:beta0_constraint}
\end{equation}
where $g$ is the gluon-quark coupling. 
$\beta_0$ can be
determined from, for example, the total $pp$ cross
section~\cite{landshoff:84,landshoff:85}.
 
As a consequence, the LN model yields simple formulae for $pp$ scattering,  
exclusive $\rho$ production in deep inelastic scattering and the $J/ \Psi - $ 
nucleon total cross section, which all contain integrals in momentum space of 
$g^2 D(p)$~\cite{halzen:93,ducati:93}. 
In order to extract predictions from the model, one needs an expression 
for $g^2 D(p)$ which is valid both in the low momentum region $p \approx$ 
1 GeV, where in fact the dominant contribution to the pomeron is 
expected, and for higher momenta, where the perturbative behaviour is 
recovered. Obviously, convergence of \eqn{eq:beta0_constraint} requires 
that the infra-red pole of the gluon propagator be removed by some 
nonperturbative mechanism. 
 
In order to use the lattice gluon propagator $D_{\rm lat} (p)$ 
in the LN model we make the 
following assumptions: 
\begin{enumerate}
\item we neglect the running of the QCD coupling, i.e. 
we make the approximation $g(p) = g$;
\item we assume that in the continuum limit   
$D_{\rm lat} (p)$ 
is multiplicatively renormalisable, as it is in perturbation theory. 
\end{enumerate}
As the scale for the momenta in $D_{\rm lat} (p)$ is set
from string tension  measurements, our assumptions imply that we only have one
free parameter   to fix in the expression $g^2 D_{\rm lat} (p)$. 
This is a multiplicative factor, that 
corresponds to the product of a gluon wavefunction renormalisation constant 
times a numerical value for $g^2$. 
We will call this parameter $g_{\rm eff}^2$, although strictly 
speaking it is more than just a squared coupling constant.

$g_{\rm eff}^2$ can be determined by using \eqn{eq:beta0_constraint} as a 
normalisation condition, i.e. by imposing that $\beta_0$ attains 
its phenomenological value of 2.0 ${\rm GeV}^{-1}$:
\begin{equation}
\beta_{0}^2 = \frac{1}{36 \pi^2} \int d^2 p \left[ g_{\rm eff}^2 \ D_{\rm lat} (p)
\right]^2 = 4 \ {\rm GeV}^{-2}
\label{eq:beta0_fix}
\end{equation}
It is the combination $g_{\rm eff}^2 D_{\rm lat} (p)$ that we insert in the
formulae of the LN model.

\begin{figure}
\begin{center}
\leavevmode
\epsfysize=250pt
\epsfbox[20 30 620 600]{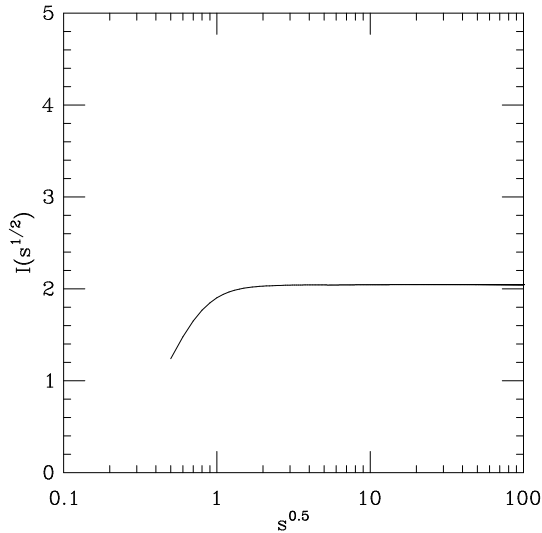}
\end{center}
\caption{The integral $I(s^{1/2})$ against $s^{1/2}$ on the $16^4$
lattice at $\beta=6.0$}
\label{fig:beta0_determination}
\end{figure}

As we mentioned in the previous section, one has to remember that the
momentum dependence of the lattice gluon propagator is influenced by
lattice artifacts. At high momentum, when $p$ becomes comparable to
the ultraviolet cutoff $a^{-1}$, the propagator is affected by
discretisation errors.  In order to judge their likely importance in
this calculation, we show in \fig{fig:beta0_determination} the
integral
\begin{equation}
I(s^{1/2}) = \frac{1}{36 \pi^2} \int_0^{s^{1/2}} 
d^2 p \left[D_{\rm lat}(p)\right]^2,
\label{eq:beta0_integ}
\end{equation}
that we use to constrain $g_{\rm eff}$, for the $16^4$ lattice at
$\beta = 6.0$.  The dominant contribution to the integral indeed
arises from momenta with $|pa| < 1$, suggesting that our evaluation of
the integral \eqn{eq:beta0_constraint} should suffer only weakly from
discretisation uncertainties.  A similar conclusion can be drawn from
the other lattices in this study.

For low momenta, finite volume effects are present, making it
difficult to identify a possible gluon mass.  In order to isolate the
extent to which such effects contaminate our calculation, we need a
comparison with a calculation at a different lattice volume, but at
the same lattice spacing.  Such a comparison is provided by
\refer{stella:94}, where a high-statistics measurement of the gluon
propagator is made on a $24^3 \times 32$ lattice at $\beta = 6.0$.  In
the remainder of this paper, we will quote results using this
determination of the lattice propagator, as well as our own
determination.

The inverse ``radius'' of the Pomeron, $\mu_0$, can be determined from
the study of exclusive $\rho$-production in Deep Inelastic Scattering.
It is related to the non-perturbative gluon propagator by
\begin{equation}
\mu_0^2 = \frac{\int d^2 p \, p^2 \left[ g^2 \, D(p) \right]^2}
{\int d^2 p \, \left[ g^2 \, D(p) \right]^2}, \label{eq:mu0}
\end{equation}
and thus is independent of the coupling $g$ in our approximation, 
and furthermore should
not depend on how we choose to parameterise the measured lattice gluon
propagator.  The phenomenological value is $\mu_0 = 1.1 \, \gev$.  The
determination of $\mu_0$ on each lattice is shown in \tab{tab:geff};
the consistency between the different measurements is encouraging, and
the results close to the phenomenological value.
\begin{table}
\begin{center}
\begin{tabular}{|c|c|c|c|}\hline
&$\beta$ & Size & $\mu_0^2\,(\gev^{2})$ \\ \hline
This work & 6.0 & $16^4$ & 0.85 \\ \hline
This work & 6.2 & $16^4$ & 1.03 \\ \hline
\refer{stella:94} & 6.0 & $24^3 \times 32$ & 0.93 \\\hline
\end{tabular}
\caption{The value of $\mu_0^2$ defined by \eqn{eq:mu0} on each lattice.}
\label{tab:geff}
\end{center}
\end{table}

We now have the ingredients to explore the phenomenology of the 2GE
model: the $p^2$ dependence of the propagator determined from a
lattice calculation, and an effective gluon-quark coupling determined
from the fit to $\beta_0$.  We begin the exploration with the analysis
of elastic proton-proton scattering.

\subsection{Proton-proton elastic scattering}

The total and elastic differential cross section for proton-proton
scattering provides a benchmark for the 2GE model of the Pomeron.
Indeed it has already been explored extensively using nonperturbative
propagators obtained from the approximate solution of the
Schwinger-Dyson equation~\cite{cudell:91,halzen:93}.  We review the
formalism here, and then apply the model using the lattice gluon
propagator.

The 2GE amplitude for elastic proton-proton scattering can be
written~\cite{gunion:77, levin:81}
\begin{equation}
A(s,t) = i s 8 \alpha_s^2 \left[ T_1 - T_2 \right]
\end{equation}
with
\begin{eqnarray}
T_1 & = & \int_0^s d^2 k D(q/2 + k) D(q/2-k) 
\left[ G_p(q,0) \right]^2 \\
T_2 & = & \int_0^s d^2 k D(q/2 + k) D(q/2 - k) G_p(q,k-q/2) \nonumber \\
& & \left[2 G_p(q,0) - G_p(q,k-q/2)\right]
\end{eqnarray}
where $G_p(q,k)$ represents a convolution of proton wave functions.
Here $T_1$ and $T_2$ represent the contributions when both gluons
attach to the same quark and to different quarks within the proton
respectively; our expectation~\cite{ln:87}, motivated by the additive
quark rule, is that $T_1$ should dominate $T_2$.  Following
ref.~\cite{cudell:91}, we take $G_p(q,k-q/2) = F_1(q^2 +
|k^2-q^2/4|)$, where $F_1$ is the elastic isoscalar form factor of the
proton,
\begin{equation}
F_1(t) = \frac{4m^2 - 2.79t}{(4m^2-t)(1-t/0.71)^2}.\label{eq:isoscalar}
\end{equation}
The total and elastic differential cross-sections are given by
\begin{eqnarray}
\sigma^0_{\rm tot} & = & \frac{A(0)}{is} \\
\frac{d \sigma^0}{dt} & = & \frac{|A(t)|^2}{16 \pi s^2}
\end{eqnarray}
respectively. Single Pomeron exchange is expected to dominate the
differential cross-section out to $-t \simeq 0.5 \, \gev^2$.  To fully
describe the energy dependence, the intercept of the Pomeron
trajectory is taken to be somewhat larger than 1, and the measured
total and differential cross sections are related to the energy-independent
expressions above by
\begin{eqnarray}
\sigma_{\rm tot} & = & \left(\frac{s}{m_p^2}\right)^{0.08}
\sigma^0_{\rm tot} \\
\frac{d \sigma}{dt} & = & \left(\frac{s}{m_p^2}\right)^{0.168}
\frac{d \sigma^0}{dt}. \label{eq:dsig_energy}
\end{eqnarray}
For small $t$, the elastic differential cross section behaves like
$e^{B t}$, and the model is characterised by two parameters,
$\sigma^0_{\rm tot}$ and $B$.

Following the discussion in the previous subsection, we compute
$\sigma^0_{\rm tot}$ and $B$ on each lattice using the lattice
gluon propagator and the effective coupling $g_{\rm eff}$.
Our results are presented in
Table~\ref{tab:pp_cross_section}.
\begin{table}
\begin{center}
\begin{tabular}{|c||c|c|c|c|}\hline
& $\beta$ & Size & $\sigma^0_{\rm tot}\,({\rm mb})$ & 
$B\,(\gev^{-2})$\\ \hline
This work & 6.0 & $16^4$ & 18.12 &  13.6 \\ \hline
This work & 6.2 & $16^4$ & 19.21 &  12.9 \\ \hline
\refer{stella:94} & 6.0 & $24^3 \times32 $ & 19.85 & 12.6 \\ \hline
\end{tabular}
\caption{The energy-independent total cross section, $\sigma^0_{\rm
tot}$, and the logarithmic slope parameter, $B$.}
\label{tab:pp_cross_section}
\end{center}
\end{table}
The general agreement between the values of $\sigma^0_{\rm tot}$ and
$B$ obtained on the various lattices is striking, suggesting either
that both quantities are subject to only small discretisation or
finite volume effects, or that such effects are absorbed in $g_{\rm
eff}$.  They are also encouragingly close to the phenomenological
values of $\sigma^0_{\rm tot} \simeq 22.7\,{\rm mb}$ and $B \sim
11~\gev^{-2}$.

\begin{figure}
\begin{center}
\leavevmode
\epsfysize=250pt
\epsfbox[20 30 620 600]{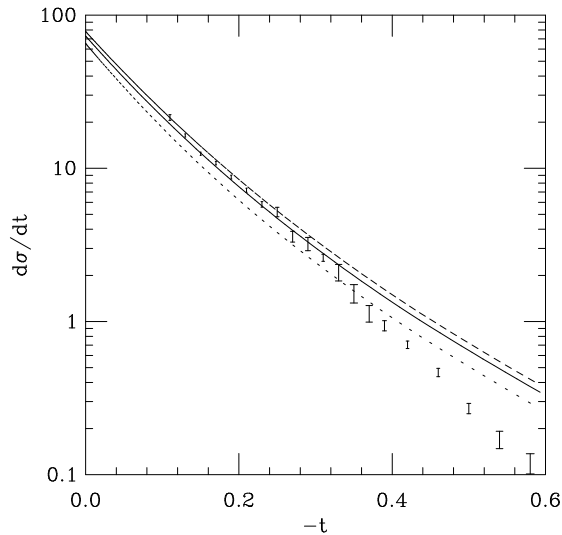}
\end{center}
\caption{Data for the $pp$ elastic cross section at $\protect\sqrt{s} =
53~\gev$ from Ref.~\protect\cite{isr:84}, together with the lattice
prediction corrected for energy dependence on the $16^4$ lattices at
$\beta = 6.2$ (solid) and $\beta = 6.0$ (dots), and on the $24^3
\times 32$ lattice at $\beta=6.0$ (dashes).\label{fig:isr_data}}
\end{figure}
The quality of the fit to the data can be seen in
\fig{fig:isr_data}, where we show ISR data for the differential
elastic cross section at $\sqrt{s} = 53~\gev$ together with the
lattice predictions, with the energy correction of
\eqn{eq:dsig_energy}.  Though the agreement of the lattice computation
with the experimental data is good, differences between the
lattice results are now clearer, and suggest the need to repeat the
calculation at different volumes and at different lattice spacings.
In particular, our lattice at $\beta=6.2$ corresponds to a somewhat tiny 
physical volume for hadronic physics, so that finite volume effects may be 
non-negligible in that case.

\subsection{$J/\psi$-Nucleon Scattering}
This process provides a further important test of the 2GE model, and
in the following we adopt the analysis procedure of Ducati \textit{et
al.}~\cite{ducati:93}.  The amplitude for meson-nucleon scattering is
\begin{eqnarray}
\lefteqn{A(s,t) = is\frac{32}{9} \alpha_s^2 \int_0^s d^2k 
D(k) D(2q-k)}\nonumber \\
& & 2\left[ f_M(q^2) - f_M((q-k)^2)\right]
3\left[f_N(q^2) - f_N(q^2 - \frac{3}{2} \vec{q}\cdot\vec{k} +
\frac{3}{4}k^2)\right].\label{eq:jpsinucleon}
\end{eqnarray}
We use the pole approximation for the
form factors
\begin{equation}
f_i(k^2) = \frac{1}{(1 - \frac{1}{6}k^2
\langle r_i^2\rangle)},\label{eq:meson_ff}
\end{equation}
taking~\cite{povh:87} $\langle r_p^2\rangle = 0.67\, \fm^2$,
$\langle r_\pi^2 \rangle = 0.44\,\fm^2$, $\langle r_K^2 \rangle = 0.35\,
\fm^2$ and $\langle r_{J/\psi}^2 \rangle = 0.04\, \fm^2$.

\begin{figure}
\begin{center}
\leavevmode
\epsfysize=250pt
\epsfbox[20 30 620 600]{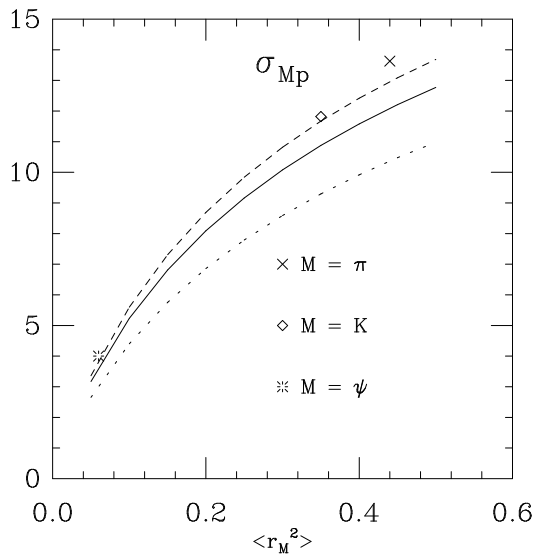}
\end{center}
\caption{The meson-nucleon total cross section as a function of the
radius used in the meson form factor, \protect\eqn{eq:meson_ff}, on
the $16^4$ lattices at $\beta=6.2$ (solid) and $\beta=6.0$ (dots), and
on the $24^3 \times 32$ lattice at $\beta=6.0$ (dashes).  Also shown
are the radii and total cross sections corresponding to the $\pi$, $K$
and $J/\psi$.\label{fig:jpsi}}
\end{figure}
Two phenomenological features are evident in the calculation.  Firstly, the
quark-counting rule is closely satisfied in the case of the hadrons
composed of light quarks:
\begin{equation}
\begin{array}{lrr}
\mbox{$\beta = 6.0$,}& \mbox{$16^4$:} &
\sigma_{pp}/\sigma_{\pi p} = 1.8, \\ 
\mbox{$\beta = 6.2$,}& \mbox{$16^4$:} &
\sigma_{pp}/\sigma_{\pi p} = 1.6, \\ 
\mbox{$\beta = 6.0$,}& \mbox{$24^3 \times 32$:} & 
\sigma_{pp}/\sigma_{\pi p} = 1.5.
\end{array}
\end{equation}
Secondly, the Pomeron couples more weakly to mesons composed of
heavier quarks.  This can be seen in
\fig{fig:jpsi} where we show the meson-nucleon cross section as a function
of the pole radius used in \eqn{eq:meson_ff}, together with a Regge
fit to the energy-independent part of the $\pi^{-} p$ and $K^{-} p$
cross sections performed in \refer{landshoff:92}.  The differences
between the results on the different lattices reinforces the need to
repeat the calculation on a wider range of lattice parameters.

\section{Gauge Dependence}

The issue of gauge dependence in the LN model is a difficult one.
Indeed, the starting point of LN is the leading-order perturbative
calculation of a quark-quark scattering amplitude, with the constraint
that no colour is exchanged between quarks. This is of course a gauge
invariant quantity, expressed by a two-gluon exchange diagram.  On the
other hand, LN argue that the salient features of pomeron physics can
be captured by substituting in such a diagram the tree-level gluon
propagator with an effective one, including non-perturbative
self-energy effects, which should in particular account for the
removal of the infrared pole. Clearly any {\it ansatz} for the
self-energy introduces in principle an uncontrollable gauge dependence
in the calculation.  Given the phenomenological success of the model,
one is led to think that maybe the gauge dependence intrinsic in the
propagator is either mild by itself or gets suppressed (cancels) in
the evaluation of physical quantities.  We emphasise that this problem
is inherent in the model and has nothing to do with using a lattice
propagator.  From the point of view of our method, one may speculate
that the simplest conceivable mechanism for the suppression of gauge
dependence would be if it gets ``factored out" in the definition of
$g_{\rm eff}$. In other words, a gauge dependent definition of $g_{\rm
eff}$ may cancel most of the gauge dependence in the momentum space
integrals of the propagator.

We plan to investigate directly the issue of gauge dependence in a
future work by using lattice propagators obtained in different gauges.
 
\section{Conclusions}
We have demonstrated how the Landau-gauge lattice gluon propagator
employed in the two-gluon-exchange model of proton-proton scattering
provides a highly successful description of the data, where the only
parameter that needs to be fixed is an effective quark-gluon coupling.
For the first time a genuine QCD quantity, evaluated entirely from first 
principles, has been inserted in the LN model, providing an important 
consistency check of the model itself. Also, given the fact that in our
approximation the effect of quark loops diagrams is absent, our analysis 
shows the robustness of the LN model, despite its simplicity.

Although the description of the data appears largely independent both of the
volume of the lattice and of the lattice spacing, a more detailed analysis  of
lattice systematic errors is necessary. We plan to perform such an analysis, 
together with an investigation of the gauge dependence of our results, in a 
future publication. 

\section*{Acknowledgements}
We wish to thank Sandy Donnachie and Peter Landshoff for useful
conversations, and Chris Michael both for helpful conversations and
for reading the manuscript.  This work was supported by the United
Kingdom Particle Physics and Astronomy Research Council (PPARC) under
grant GR/J21347.  CP and DGR acknowledge the support of PPARC through
Advanced Fellowships held at the University of Liverpool and at the
University of Edinburgh respectively.  We are grateful to Edinburgh's
Computing Service for maintenance of service on the Thinking Machines
CM-200 where this work was performed.

\end{document}